\documentclass{emulateapj}
\pdfoutput=1
\usepackage{graphicx}
\usepackage{amsmath}
\usepackage{graphicx}
\usepackage{l3regex}
\usepackage{mhchem}
 \newcounter{chem}
\newcounter{temp}

\newenvironment{chequation}{%
  \setcounter{temp}{\value{equation}}%
  \setcounter{equation}{\value{chem}}%
  \renewcommand{\theequation}{R\arabic{equation}}%
}{%
  \setcounter{chem}{\value{equation}}%
  \setcounter{equation}{\value{temp}}%
}
 
\begin{document}

\shorttitle{The K Dwarf Advantage for Biosignatures}
% If more than two authors, use {\em et al.}
\shortauthors{Arney}

\title{{\bf The K Dwarf Advantage for Biosignatures on Directly Imaged Exoplanets}}
%% AUTHOR/INSTITUTIONS FOR AASTEX6.1:

\author{Giada~N.~Arney}
\affiliation{Planetary Systems Laboratory, NASA Goddard Space Flight Center, Greenbelt, MD 20771, USA}
\affiliation{NASA NExSS Virtual Planetary Laboratory, P.O. Box 351580, Seattle, WA 98195, USA}
\affiliation{Sellers Exoplanet Environments Collaboration, NASA Goddard Space Flight Center, Greenbelt, MD 20771, USA}

%% AUTHOR/INSTITUTIONS FOR EMULATE APJ:
% \author{Patricio~E.~Cubillos\altaffilmark{1,2},
% Joseph~Harrington\altaffilmark{1},
% and
% Third~Author\altaffilmark{1}
% }
% \affil{\sp{1} Planetary Sciences Group, Department of
%               Physics, University of Central Florida, Orlando, FL 32816-2385\\
%        \sp{2} Space Research Institute, Austrian Academy of Sciences,
%               Schmiedlstrasse 6, A-8042, Graz, Austria}

\email{giada.n.arney@nasa.gov}

% %% Extra info for aastex:
% \received{Yesterday}
% \revised{Today}
% \accepted{Tonight}
% \published{Tomorrow}
% \submitjournal{AASJournal}

\begin{abstract}
Oxygen and methane are considered to be the canonical biosignatures of modern Earth, and the simultaneous detection of these gases in a planetary atmosphere is an especially strong biosignature. However, these gases may be challenging to detect together in the planetary atmospheres because photochemical oxygen radicals destroy methane. Previous work has shown that the photochemical lifetime of methane in oxygenated atmospheres is longer around M dwarfs, but M dwarf planet habitability may be hindered by extreme stellar activity and evolution. Here, we use a 1-D photochemical-climate model to show that K dwarf stars also offer a longer photochemical lifetime of methane in the presence of oxygen compared to G dwarfs. For example, we show that a planet orbiting a K6V star can support about an order of magnitude more methane in its atmosphere compared to an equivalent planet orbiting a G2V star. In the reflected light spectra of worlds orbiting K dwarf stars, strong oxygen and methane features could be observed at visible and near-infrared wavelengths. Because K dwarfs are dimmer than G dwarfs, they offer a better planet-star contrast ratio, enhancing the signal-to-noise (SNR) possible in a given observation. For instance, a 50 hour observation of a planet at 7 pc with a 15-m telescope yields SNR = 9.2 near 1 $\mu$m for a planet orbiting a solar-type G2V star, and SNR = 20  for the same planet orbiting a K6V star. In particular, nearby mid-late K dwarfs such as 61 Cyg A/B, Epsilon Indi, Groombridge 1618, and HD 156026 may be  excellent targets for future biosignature searches. 
\end{abstract}

% http://journals.aas.org/authors/keywords2013.html

\section{INTRODUCTION}
\label{introduction}
One of the most profound scientific questions that could be answered in the near future is whether there is life on other planets. Future telescopes will seek remotely detectable signs of life, or biosignatures, in exoplanet atmospheres.  The most studied approach to biosignature search strategies is detection of an atmosphere in chemical disequilibrium \citep[e.g.,][]{Lovelock1965, Hitchcock1967, Sagan1993, Kaltenegger2007, Krissansen-Totton2016, Schwieterman2018, Krissansen-Totton2018}. For modern Earth, the largest overall disequilibrium is caused by the simultaneous presence of oxygen (\ce{O2}), atmospheric nitrogen (\ce{N2}), and liquid water (\ce{H2O}), which would react to form nitrate and hydrogen ions in equilibrium \citep{Krissansen-Totton2016}. Unfortunately, \ce{N2} may be challenging to observe in direct spectral observations \citep{Schwieterman2015a}, so other directly delectable biosignatures should be sought. 

The ``canonical'' biosignature disequilibrium pairing for modern Earth is the simultaneous presence of \ce{O2} and methane (\ce{CH4}), whose atmospheric abundances are orders of magnitude away from equilibrium values \citep{Lovelock1965, Hitchcock1967}. These gases, which both produce spectral features at visible and near-infrared (NIR) wavelengths, will be high priority gases sought in future biosignature searches. However, despite their importance as biosigantures, oxygen and methane have not always been delectable in Earth's atmosphere over our planet's geological history. Briefly, we will review the history of oxygen and methane in Earth's atmosphere in order to provide context and motivation for the search for these gases on exoplanets. 

\begin{figure*}[hbt]
\begin{center}
\includegraphics[width=152mm,scale=1.0]{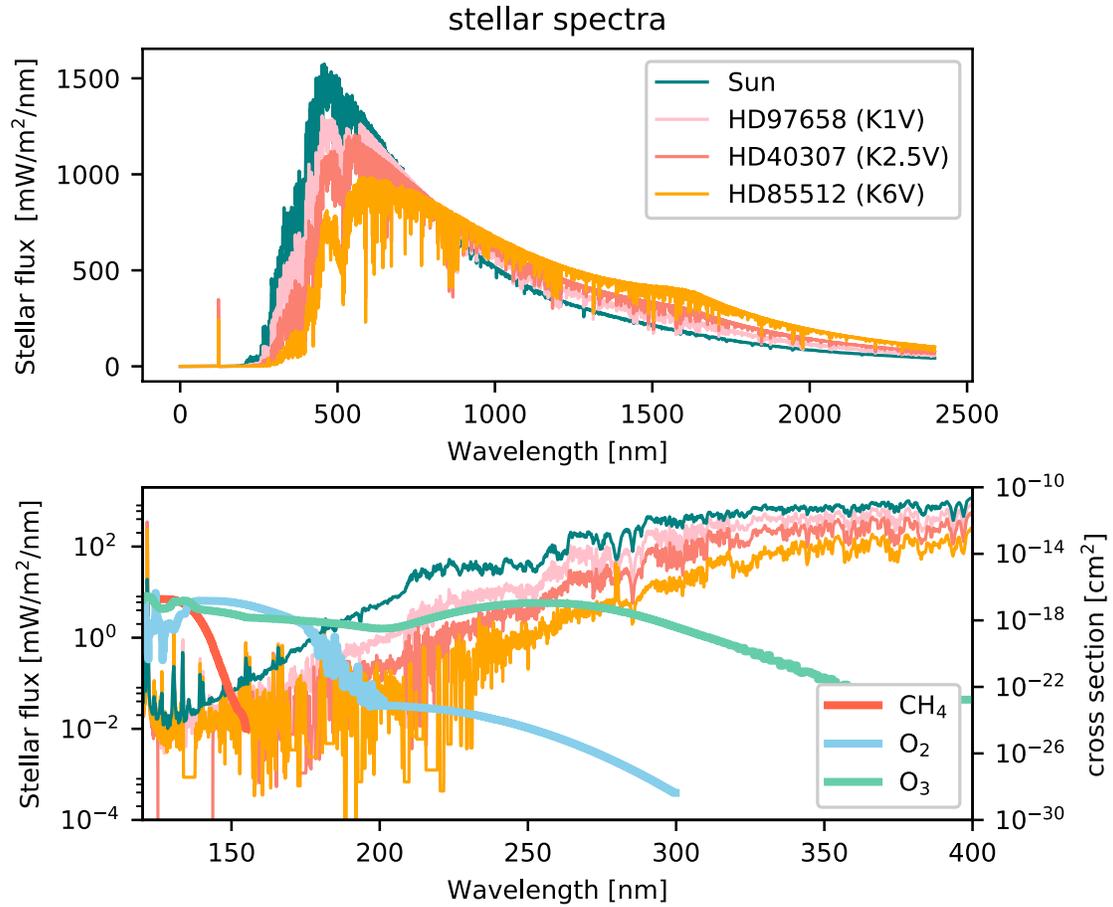}
\caption{\emph{Top panel:} Stellar spectra used in our simulations. \emph{Bottom panel:} The stellar UV wavelengths. Also shown in the bottom panel are UV cross sections of \ce{CH4}, \ce{O2}, and \ce{O3} (right y-axis).} 
\end{center}
\label{fig:fig1}
\end{figure*}

Oxygenic photosynthesis, the dominant metabolism on our planet today, probably evolved on Earth by 3 billion years ago (Ga, short for ``giga-annum'') \citep{Planavsky2014b}, and potentially as early as 3.7 Ga \citep{Rosing2004}. Because it uses cosmically ubiquitous compounds (\ce{H2O}, \ce{CO2}, starlight) and has a high energy yield, this metabolism may be incentivized to evolve elsewhere. 

However, atmospheric oxygenation depends not only \ce{O2} production, but also on the competition between \ce{O2} sources and sinks. While oxygenic photosynthesis likely existed earlier, it was not until roughly 2.3 Ga (the start of the Proterozoic geological eon) that atmospheric oxygen levels rose during the ``Great Oxygenation Event.'' However, oxygen levels were likely variable throughout the Proterozoic, and in the mid-Proterozoic (2.0-0.7 Ga), atmospheric oxygen abundance may have been much lower than modern levels, possibly lower than 0.1$\%$ of the present atmospheric level (PAL) \citep{Planavsky2014}. At an \ce{O2} level this low, oxygen itself cannot be observed directly in the planet's spectrum, but it might be indirectly inferred by detecting its photochemical byproduct, ozone (\ce{O3}), which produces a strong spectral feature at ultraviolet (UV) wavelengths even when \ce{O2} itself is spectrally invisible. 

Methane also has a long history of biogenic production on Earth. Methanogenesis, a simple anaerobic metabolism that produces methane from \ce{CO2} and \ce{H2}, is rooted deeply in Earth's tree of life \citep{Woese1977, Ueno2006}. Methanogenesis may even have evolved as early as during the Hadean geological period \citep[prior to 3.8 Ga, a time when Earth's geological record has nearly entirely vanished; ][]{Battistuzzi2004}. While \ce{CH4} also has geological sources, most of the \ce{CH4} in Earth's atmosphere today is biogenic \citep{Etiope2013}, and this was likely also the case for early Earth \citep{Kharecha2005}. 

Despite the continual production of biogenic \ce{O2} and \ce{CH4} on Earth for billions of years, these gases may not have produced simultaneously detectable spectral features over our planet's history \citep{Olson2016, Reinhard2017}. Methane is readily destroyed by oxygen radicals in an atmosphere containing oxygen around a G2V star. The dominant sink of \ce{CH4} on Earth starts with photodissociation of \ce{O3} (which itself forms from \ce{O2} photochemistry): 

\begin{chequation}
\begin{align}
\ce{O3 + h$\nu$ -> O2 + O^1D}  \label{ch_1}\\ 
\ce{O^1D + H2O -> 2OH} \label{ch_2}\\
\ce{CH4 + OH -> CH3 + H2O} \label{ch_3}
\end{align}
\end{chequation}

Because this mechanism is driven by photochemistry, different host stars may lead to different photochemical outcomes. \citet{Segura2005} showed that \ce{CH4} has a longer photochemical lifetime in the atmospheres of Earthlike planets orbiting M dwarf stars, which produce less radiation compared to the Sun at 200-350 nm where \ce{O3} is photolyzed. This increases the \ce{CH4} photochemical lifetime from 10 years for a planet orbiting the Sun to about 200 years. It may therefore be easier to simultaneously detect \ce{CH4} and \ce{O2} for planets around M dwarfs \citep[see also, e.g., ][]{Meadows2018}. 

Potentially habitable planets orbiting M dwarfs will likely be targeted by the James Webb Space Telescope (JWST) and future large ground-based observatories. Unfortunately, the habitability of M dwarf planets may be hindered by a number of complications including: extreme water loss during the extended super-luminous pre-main sequence phase \citep{Luger2015}, high x-ray luminosities \citep{Shkolnik2014}, and frequent energetic flares that may cause severe atmospheric loss \citep{Owen2016, Airapetian2017, Garcia-Sage2017}. 

Recently, the \textit{Exoplanet Science Strategy }Report \citep{ExoSciStrategy} recommended that NASA ``lead a large strategic direct imaging mission capable of measuring the reflected-light spectra of temperature terrestrial planets orbiting Sun-like stars.'' Planets orbiting F, G, and K dwarfs (i.e. ``Sun-like stars'') do not face the multiple challenges to habitability posed by M dwarfs, and they may therefore represent our best chance of discovering other planets similar to Earth. NASA has directed studies of two observatories for consideration in the astrophysics 2020 decadal survey that would be capable of directly observing temperate, Earth-sized exoplanets around Sun-like stars: the Large UV Optical Infrared surveyor (LUVOIR\footnote{https://asd.gsfc.nasa.gov/luvoir/}) and the Habitable Exoplanet Observatory (HabEx\footnote{https://www.jpl.nasa.gov/habex/}). 

Compared to F and G dwarfs, K dwarfs offer certain advantages as habitable planet hosts: they are more abundant than G and F dwarfs, comprising about $12\%$ of the main sequence stellar population (G dwarfs comprise about 8$\%$ while F dwarfs comprise a paltry 3$\%$); their lifetimes are longer than F and G dwarfs (17-70 billion years for K dwarfs, compared to 10 billion years for the Sun); and the planet-star contrast ratio is better for K dwarfs than for F and G dwarfs (a K2V star is only about a third as luminous as a G2V star, and a K6V star is only about a tenth as luminous), making their planets easier to observe via direct imaging. Many advantages of K dwarfs as habitable planet hosts are discussed in detail in \citet{Cuntz2016}.

Additionally, compared to M dwarfs, K dwarfs are less active, and their pre-main sequence phases are shorter \citep[< 0.1 Gyr compared to up to 1 Gyr for M dwarfs,][]{Luger2015}. Recently, \citet{Richey2019} measured the near-UV (NUV), far-UV (FUV), and X-ray evolution of K dwarf stars in moving groups aged from 10-625 Myr, finding that young planets orbiting K dwarfs are subjected to 5-50 times lower UV and X-ray fluxes compared to planets orbiting early M dwarfs, and 50-1000 times lower fluxes compared to planets orbiting late M dwarfs. \citet{Richey2019} also found that K dwarf FUV and X-ray fluxes decrease after $\sim$100 Myr, compared to $\sim$650 Myr for M dwarfs, which may have implications for early habitability and atmospheric evolution for planets around these different types of stars. The UV environment of a given host star is critical to consider when studying planetary habitability and photochemistry.

Here, we explore an additional advantage for K dwarfs: the hypothesis that like M dwarfs, K dwarf stellar UV spectra will result longer photochemical lifetimes for methane in oxygenated atmospheres. Previous photochemical modeling efforts of Earth-like planets orbiting K dwarfs have explored this effect \citep{Segura2003, Rugheimer2013, Rugheimer2018}, but there are differences between these studies and ours. This new analysis using a recently upgraded photochemical model focuses explicitly on determining which parts of the \ce{CH4}/\ce{O2}/stellar spectrum parameter space produce simultaneously observable \ce{CH4} and \ce{O2} spectral features for the Sun and several K dwarfs.  We allow Earth history to bound parts of the explored parameter space. However, we also simulate atmospheres that are not representative of any period of Earth history to consider exoplanets with different evolutionary paths. In addition, we discuss the implications of these results in the context of possible future exoplanet observatories, and we consider which of the nearby K dwarf stars may be the best targets for future biosignature searches. By understanding the ``K dwarf advantage,'' we improve our chances of selecting the best targets for biosignature searches with future observatories.

\section{METHODS}
\label{sec:methods}
To simulate our atmospheres, we use a coupled 1D photochemical-climate model called Atmos, which is described in \citet{Arney2016}. Its photochemical module is based on a photochemical code originally developed by \citet{Kasting1979} and significantly updated and modernized as described in \citet{Zahnle2006}. The photochemical module has recently been updated as described in \citet{Lincowski2018} with an expanded and higher resolution wavelength grid; and updated cross sections, quantum yields, and reaction rates. Tests comparing the upgraded model used here to the previous version suggest that the previous version may overestimate \ce{CH4} abundances for the types of atmospheres that we simulate here by up to 50$\%$. This upgraded model has been validated on Earth and Venus as described in \citet{Lincowski2018}. The climate module of Atmos was originally developed by \citet{Kasting1986}, and like the photochemical model has evolved considerably since this first incarnation. This climate module has recently been used to study, e.g.,  habitable zone (HZ) boundaries \citep{Kopparapu2013}. 

We simulate planets orbiting the K dwarf stars in the MUSCLES treasury survey \citep{France2016, Youngblood2016, Loyd2016}: HD 97658 (K1V), HD 40307 (K2.5V), and HD 85512 (K6V). We also include the Sun \citep{Chance2010} for comparison. These spectra and stellar properties are provided in Figure \ref{fig:fig1} and the top rows of Table \ref{tab:tab1}.

\begin{table*}[htb]
\caption{Key stellar properties and resulting planetary atmospheric properties. Note the planet for the Earth-Sun system is assumed to be at 1.2 AU because we assume all planets receive 0.7 $\times$ Earth's insolation to place all of the K dwarf planets in the habitable zone.}
\centering
\begin{tabular}{lcccc}
\hline
\hline
& Sun (G2V)    & HD 97658 (K1V) & HD 40307 (K2.5V) & HD 85512 (K6V)   \\     
\hline
Temperature (K)   & 5778                         & 4991                         & 4977                         & 4759                         \\
Mass (M$_{Sun}$)                                     & 1                            & 0.75                          & 0.75                         & 0.69                         \\
Luminosity (L$_{Sun}$)                                     & 1                            & 0.3                          & 0.23                         & 0.13                         \\
Planet-star separation (AU)                     & 1.2                          & 0.65                         & 0.57                         & 0.42                         \\ \\

\begin{tabular}[c]{@{}l@{}}\ce{CH4} surface mixing ratio \\ Case 1, 2, 3\end{tabular} & \begin{tabular}[c]{@{}c@{}}2.4$\times$10$^{-5}$\\ 9.0$\times$10$^{-6}$\\ 8.7$\times$10$^{-5}$\end{tabular}   & \begin{tabular}[c]{@{}c@{}}3.7$\times$10$^{-5}$\\ 1.3$\times$10$^{-5}$\\ 1.3$\times$10$^{-4}$\end{tabular}  & \begin{tabular}[c]{@{}c@{}}7.0$\times$10$^{-5}$\\ 2.2$\times$10$^{-5}$\\ 2.5$\times$10$^{-4}$\end{tabular}  & \begin{tabular}[c]{@{}c@{}}2.2$\times$10$^{-4}$\\ 5.0$\times$10$^{-5}$\\ 8.1$\times$10$^{-4}$\end{tabular} \\ \\ 
\begin{tabular}[c]{@{}l@{}}\ce{CH4} photolysis rate (s$^{-1}$)\\ Case 1, 2, 3\end{tabular} & \begin{tabular}[c]{@{}c@{}}3.7$\times$10$^{9}$\\ 3.0$\times$10$^{8}$\\ 9.1$\times$10$^{8}$\end{tabular}      & \begin{tabular}[c]{@{}c@{}}1.4$\times$10$^{10}$\\ 1.4$\times$10$^{9}$\\ 8.3$\times$10$^{9}$\end{tabular}    & \begin{tabular}[c]{@{}c@{}}2.8$\times$10$^{10}$\\ 3.1$\times$10$^{9}$\\ 2.3$\times$10$^{10}$\end{tabular}   & \begin{tabular}[c]{@{}c@{}}7.0$\times$10$^{10}$\\ 6.6$\times$10$^{9}$\\ 6.8$\times$10$^{10}$\end{tabular}  \\ \\
\begin{tabular}[c]{@{}l@{}}\ce{CH4}  column density\\ Case 1, 2, 3\end{tabular}       & \begin{tabular}[c]{@{}c@{}}4.9$\times$10$^{20}$\\ 1.8$\times$10$^{20}$\\ 1.8$\times$10$^{21}$\end{tabular}   & \begin{tabular}[c]{@{}c@{}}7.6$\times$10$^{20}$\\ 2.7$\times$10$^{20}$\\ 2.6$\times$10$^{21}$\end{tabular}  & \begin{tabular}[c]{@{}c@{}}1.4$\times$10$^{21}$\\ 4.6$\times$10$^{20}$\\ 5.0$\times$10$^{21}$\end{tabular}  & \begin{tabular}[c]{@{}c@{}}4.7$\times$10$^{21}$\\ 9.4$\times$10$^{20}$\\ 1.7$\times$10$^{22}$\end{tabular} \\ \\
\begin{tabular}[c]{@{}l@{}}\ce{O2} photolysis rate (s$^{-1}$)\\ Case 1, 2, 3\end{tabular}  & \begin{tabular}[c]{@{}c@{}}1.3$\times$10$^{12}$ \\ 2.3$\times$10$^{12}$ \\ 3.6$\times$10$^{12}$\end{tabular} & \begin{tabular}[c]{@{}c@{}}4.7$\times$10$^{11}$ \\ 9.2$\times$10$^{11}$\\ 1.2$\times$10$^{12}$\end{tabular} & \begin{tabular}[c]{@{}c@{}}3.1$\times$10$^{11}$ \\ 6.9$\times$10$^{11}$\\ 8.0$\times$10$^{11}$\end{tabular} & \begin{tabular}[c]{@{}c@{}}1.8$\times$10$^{11}$\\ 4.8$\times$10$^{11}$\\ 4.8$\times$10$^{11}$\end{tabular} \\ \\
\begin{tabular}[c]{@{}l@{}}\ce{O2} surface flux (molec/s/cm$^2$)\\ Case 1, 2, 3\end{tabular}  & \begin{tabular}[c]{@{}c@{}}7.7$\times$10$^{11}$ \\ 5.4$\times$10$^{11}$ \\ 2.8$\times$10$^{12}$\end{tabular} & \begin{tabular}[c]{@{}c@{}}7.8$\times$10$^{11}$ \\ 5.4$\times$10$^{11}$\\ 2.8$\times$10$^{12}$\end{tabular} & \begin{tabular}[c]{@{}c@{}}7.9$\times$10$^{11}$ \\ 5.4$\times$10$^{11}$\\ 2.9$\times$10$^{12}$\end{tabular} & \begin{tabular}[c]{@{}c@{}}7.8$\times$10$^{11}$\\ 5.5$\times$10$^{11}$\\ 2.9$\times$10$^{12}$\end{tabular} \\ \\
\begin{tabular}[c]{@{}l@{}}\ce{O3} photolysis rate (s$^{-1}$)\\ Case 1, 2, 3\end{tabular}  & \begin{tabular}[c]{@{}c@{}}1.3$\times$10$^{15}$\\ 1.9$\times$10$^{15}$\\ 1.5$\times$10$^{15}$\end{tabular}   & \begin{tabular}[c]{@{}c@{}}4.3$\times$10$^{14}$\\ 1.0$\times$10$^{15}$\\ 6.9$\times$10$^{14}$\end{tabular}  & \begin{tabular}[c]{@{}c@{}}1.3$\times$10$^{14}$\\ 4.9$\times$10$^{14}$\\ 2.7$\times$10$^{14}$\end{tabular}  & \begin{tabular}[c]{@{}c@{}}2.2$\times$10$^{13}$\\ 1.7$\times$10$^{14}$\\ 5.9$\times$10$^{13}$\end{tabular} \\ \\
\begin{tabular}[c]{@{}l@{}}\ce{O3} column density\\ Case 1, 2, 3\end{tabular}     & \begin{tabular}[c]{@{}c@{}}3.3$\times$10$^{18}$\\ 7.2$\times$10$^{18}$\\ 5.2$\times$10$^{18}$\end{tabular}   & \begin{tabular}[c]{@{}c@{}}1.5$\times$10$^{18}$\\ 4.7$\times$10$^{18}$\\ 2.9$\times$10$^{18}$\end{tabular}  & \begin{tabular}[c]{@{}c@{}}5.1$\times$10$^{17}$\\ 2.7$\times$10$^{18}$\\ 1.3$\times$10$^{18}$\end{tabular}  & \begin{tabular}[c]{@{}c@{}}1.3$\times$10$^{17}$\\ 1.3$\times$10$^{18}$\\ 4.0$\times$10$^{17}$\end{tabular} \\ \\

\end{tabular}
\label{tab:tab1}
\end{table*}

A planet receiving the same total insolation (i.e. incident solar energy, S$_o$) that modern Earth receives from the Sun would place it inside the inner edge of the conservative HZ for a K6V star \citep{Kopparapu2013}. Therefore, to be conservative, we set the orbital distances of each planet around each of their stars to be where they receive 0.7 times the modern Earth-equivalent insolation (0.7 $\times$ S$_o$, see Table \ref{tab:tab1}). This allows each planet to sit well within its star's HZ, allowing our climate simulations to avoid extreme warming in the methane-rich atmospheres we simulate. 

For our simulated atmospheres, we vary the oxygen partial pressure (\ce{pO2}) between $10^{-3}$ and 0.21 bar (equivalent to \ce{O2} mixing ratios of $10^{-3}$ and 0.21), which brackets higher Proterozoic-like \ce{O2} levels and the modern \ce{O2} abundance. We exclude the lower mid-Proterozoic oxygen estimates \citep[p\ce{O2} $\leq$ 10$^{-4}$,][]{Planavsky2014b} because our goal is to explore the phase space of atmospheres with directly detectable oxygen and methane, but the lowest oxygen estimates for the Proterozoic do not produce directly detectable \ce{O2} spectral features.  We vary \ce{CH4} fluxes (\textit{f}\ce{CH4}) at the surface of the model between $7\times10^{10} –- 10^{12}$ molecules/cm$^2$/s. This brackets the modern methane flux to roughly an order of magnitude greater methane than is produced on modern Earth. Note that higher \ce{CH4} fluxes than the modern may be possible for Archean Earth \citep{Kharecha2005}, and this range of fluxes is sufficiently high enough that abiotic production at these rates is implausible on an exoplanet \citep{Krissansen-Totton2018}. We do not explore methane flux levels less than the modern production rate because lower fluxes do not generate strongly detectable methane spectral features for the oxygenated atmospheres that we simulate.  Our upper bound for \ce{CH4} production allows us to consider an optimistic case for an exoplanet with much more vigorous biotic methane production than Earth. 

The partial pressure of \ce{CO2} (p\ce{CO2}) is set at 0.01 bar to provide greenhouse warming in our coupled climate-photochemical model for insolation of 0.7 x S$_o$. A total surface pressure of 1 bar is assumed for all atmospheres. 

Spectra are generated using the Spectral Mapping Atmospheric Radiative Transfer Model (SMART) \citep{Meadows1996}, a 1-D line-by-line fully multiple scattering radiative transfer model. The surface albedo in our spectral model uses a composite average of 65.6$\%$ seawater, 23.1$\%$ soil/desert, and 11.3$\%$ snow/ice as described in \citet{Meadows2018}. Patchy clouds in our disk-integrated spectra are included by constructing weighted averages with 50$\%$ clear sky, 25$\%$ cirrus clouds, and 25$\%$ stratocumulus clouds. Coronagraph simulations of observations with possible future telescopes are generated with the model described in \citet{Robinson2016} with updates described in \citet{Meadows2018}. An online version of this model is available at the LUVOIR website\footnote{https://asd.gsfc.nasa.gov/luvoir/tools/}.

\section{RESULTS}
\label{sec:results}

\begin{figure*}
\begin{center}
\includegraphics[width=182mm,scale=1.0]{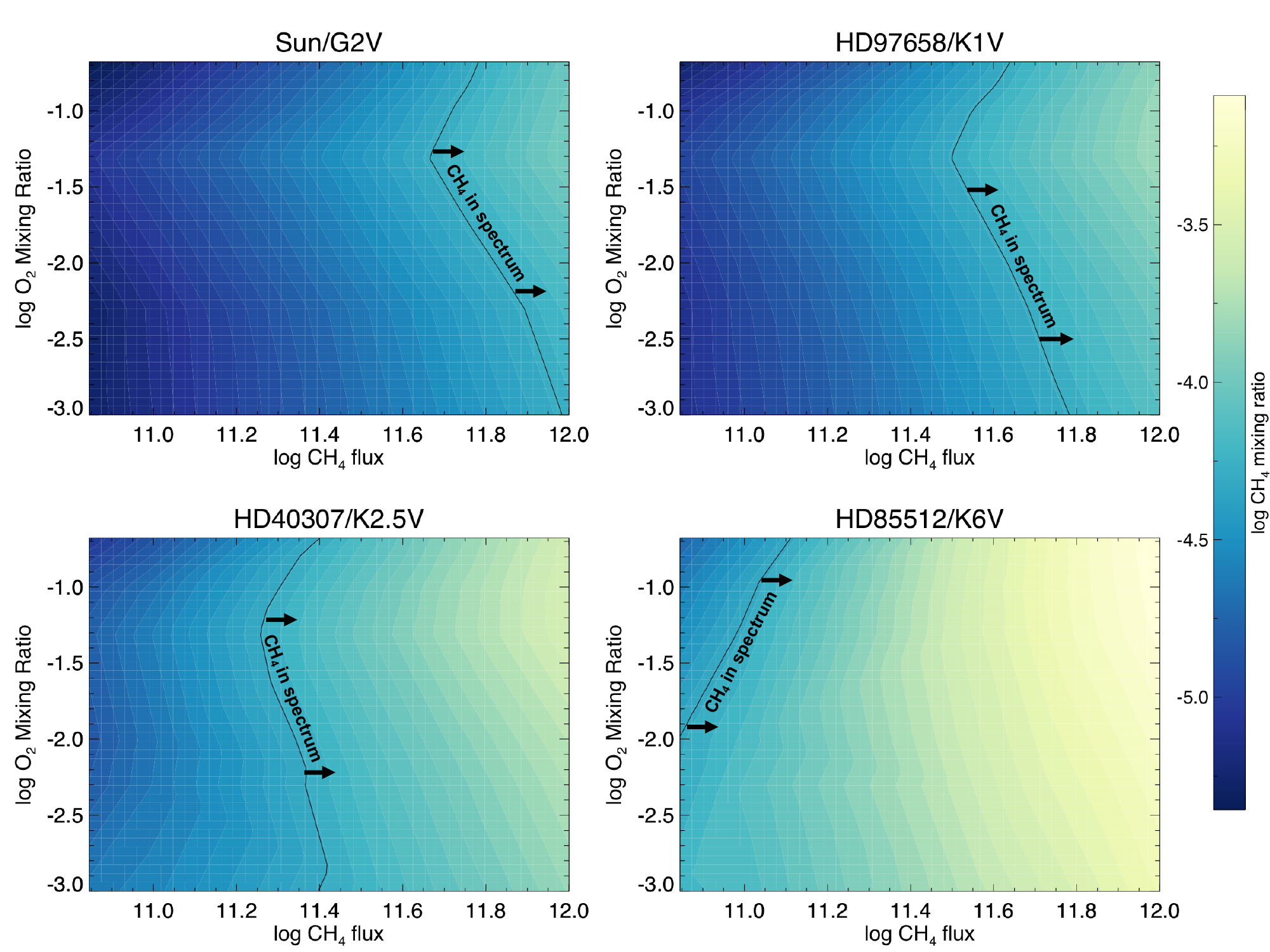}
\caption{Planetary surface methane mixing ratios as a function of stellar type, methane flux at the planet surface, and surface \ce{O2} mixing ratio. Lower-mass stars with less UV flux generate fewer photochemical oxygen radicals, which leads to more methane present in their planets' atmospheres. The solid line indicates atmospheres with methane mixing ratios of $5\times10^{-5}$, which allows for  methane features to begin to become apparent in the spectrum near 1.4 and 1.15 $\mu$m. We consider atmospheres to the right of this line especially useful candidates for methane detection in direct imaging.} 
\end{center}
\label{fig:fig2}
\end{figure*}

Figure \ref{fig:fig2} shows how later host star types lead to increasing concentrations of methane as a function of methane flux and oxygen concentration. For our analysis, we select three points from  this parameter space for each star as interesting case studies for deeper analysis. Where possible, we consider atmospheres that are consistent with the biogeochemical calculations of \citet{Olson2016}, which describes how methane fluxes to the atmosphere are impacted by p\ce{O2}. 
\begin{itemize}
\item ``Case 1'' represents a Proterozoic-like planet that is consistent with \citet{Olson2016} with p\ce{O2} = $5\times10^{-3}$ bar and \textit{f}\ce{CH4} = $3\times10^{11}$ molecules/cm$^2$/s. 
\item ``Case 2'' is a quasi-modern Earth-like planet also consistent with the calculations of \citet{Olson2016} with p\ce{O2} = 0.1 bar and \textit{f}\ce{CH4} = $1\times10^{11}$ molecules/cm$^2$/s. 
\item ``Case 3'' represents a type of exoplanet with modern Earth \ce{O2} and significantly higher \ce{CH4} production than Earth: p\ce{O2} = 0.21 bar, \textit{f}\ce{CH4} = $1\times10^{12}$ molecules/cm$^2$/s. This planet allows us to examine an optimistic case of high biological production of both of these gases on a world that is different from Earth.
\end{itemize}

\begin{figure*}[bth]
\begin{center}
\includegraphics[width=182mm,scale=1.0]{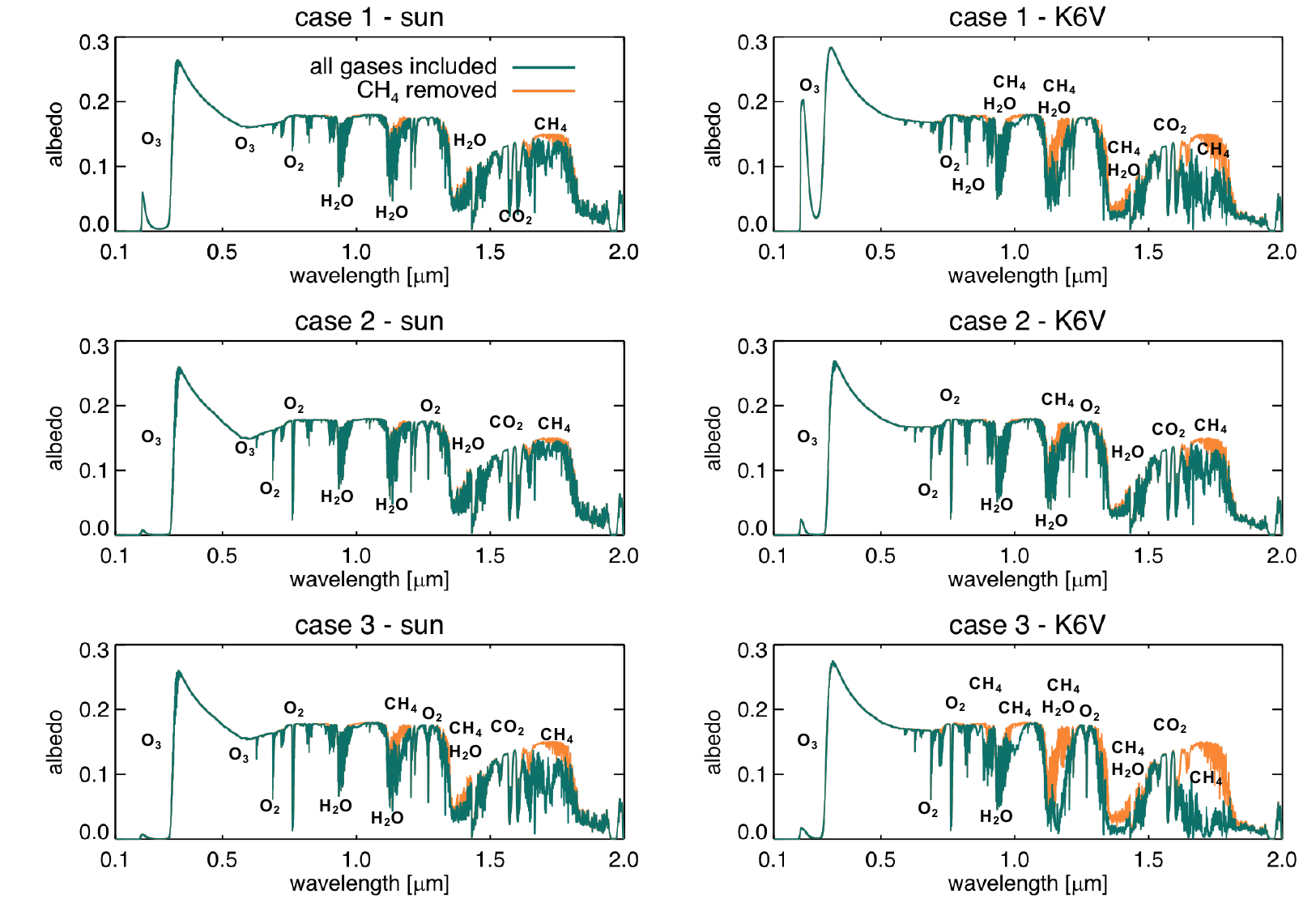}
\caption{Spectra of selected parts of parameter space for the solar-type and the K6V planets. In all cases, methane features are stronger for the planets around the K6V star. The teal colored spectra include all gases; methane has been removed from the orange spectra to so that its absorption features can be easily seen. Case 1 is the Proterozoic-like planet (p\ce{O2} = $5\times10^{-3}$ bar and \textit{f}\ce{CH4} = $3\times10^{11}$ molecules/cm$^2$/s), Case 2 is the quasi-modern planet (p\ce{O2} = 0.1 bar and \textit{f}\ce{CH4} = $1\times10^{11}$ molecules/cm$^2$/s), and Case 3 is the highest \ce{CH4}/highest \ce{O2} scenario (p\ce{O2} = 0.21 bar, \textit{f}\ce{CH4} = $1\times10^{12}$ molecules/cm$^2$/s).} 
\end{center}
\label{fig:fig3}
\end{figure*}

The K dwarfs studied here produce less radiation than the Sun at almost all UV wavelengths, so \ce{O3} is less readily photolyzed (Figure \ref{fig:fig1}, bottom panel), generating fewer oxygen radicals and allowing enhanced accumulation of \ce{CH4}. Interestingly, the K dwarf planets also generate less \ce{O3} to begin with compared to the planets orbiting the Sun because they also less readily photolyze \ce{O2}, which is needed to generate \ce{O3} (Table \ref{tab:tab1}). As shown in Table \ref{tab:tab1}, photolysis of \ce{O2} is almost an order of magnitude slower for planets orbiting the K6V star compared to  equivalent planets orbiting the Sun. As a result of these factors, the K dwarf planets have more methane than the planets orbiting the Sun, with methane levels increasing toward later type K dwarf stars. A planet orbiting the K6V star used here can have about an order of magnitude more \ce{CH4} in an oxygenated atmosphere compared to an equivalent planet around the Sun (Table \ref{tab:tab1}). Additionally, our results show a trend of increasing \ce{CH4} towards higher oxygen mixing ratios, peaking for most stars at about p\ce{O2} $= 10^{-1.3}$ bar, which is attributable to shielding of \ce{CH4} from photolysis by \ce{O2} and \ce{O3} \citep[e.g.][and note the overlap in the UV cross sections of \ce{CH4} with \ce{O2} and \ce{O3} in Figure \ref{fig:fig1}]{Olson2016}. 

While planets orbiting K dwarfs have more \ce{CH4} relative to equivalent planets around G dwarfs, M dwarfs offer the potential for even more \ce{CH4} in equivalent atmospheres becuase they produce even lower levels of radiation at the UV wavelengths needed for ozone photolysis \citep{Segura2005, Meadows2018}. For instance, \citet{Segura2005} found that methane levels increased by over two orders of magnitude for modern Earth-like planets orbiting the M dwarfs AD Leo and GJ 436C compared to planets orbiting the Sun. However, M dwarfs may be problematic habitable planet hosts for the reasons discussed in Section \ref{introduction}.

Table \ref{tab:tab1} also shows the \ce{O2} surface flux required to produce the constant mixing ratios selected from our parameter space to represent Cases 1, 2, and 3. In each of these cases, the \ce{O2} flux is controlled by the \ce{O2} mixing ratio selected and the amount of reductants in the atmosphere. In each of these cases, the \ce{O2} fluxes are larger than the \ce{CH4} surface fluxes. Biogenic methane fluxes larger than the oxygen fluxes may be difficult to sustain on planets with primary productivities driven by oxygenic photosynthesis \citep{Zerkle2012}, so these atmospheres are consistent with this constraint.  

Future direct imaging observatories being studied, such as LUVOIR and HabEx, are baselined to be able to observe exoplanets to a longest wavelength of 2 $\mu$m (at longer wavelengths, thermal radiation from the telescope swamps the planet signal).  Methane begins to become weakly apparent in the planet spectrum at 1.7 $\mu$m  for mixing ratios of about $1\times10^{-6}$ bar (this is roughly the concentration of \ce{CH4} in modern Earth's atmosphere), but detecting \ce{CH4} at this low abundance would be extremely challenging \citep[e.g.][]{Reinhard2017}. Even at higher methane abundances, observing 1.7 $\mu$m may be difficult because longer wavelengths are vulnerable to falling inside the telescope's inner working angle (IWA) (see Section \ref{sec:discussion}), especially for K dwarf planets that orbit closer to their stars than planets in the HZs of G dwarfs.  

At a methane mixing ratio of roughly $5\times10^{-5}$ bar, \ce{CH4} features near 1.4 and 1.15 $\mu$m begin to become weakly apparent in the spectrum. Therefore, we consider \ce{CH4} mixing ratios $> 5\times10^{-5}$ bar to be best for detecting methane for direct imaging observatories that can observe wavelengths $<$ 2  $\mu m$ since these wavelengths are less likely to be cut off by the IWA. We indicate this part of parameter space with the solid line on each panel in Figure \ref{fig:fig2}: atmospheres to the right of this line have \ce{CH4} mixing ratios $> 5\times10^{-5}$ bar.

Several of the atmospheres that we simulate have high enough oxygen and methane concentrations that both of these gases produce prominent spectral features at visible and NIR wavelengths accessible to direct imaging observatories. Figure \ref{fig:fig3} shows spectra for Cases 1, 2, and 3 for planets orbiting the Sun, which has the weakest \ce{CH4} features, and for planets orbiting the K6V star, which has the strongest \ce{CH4} features.  Notably, Cases 1 and 3 allow access to methane features near and shortward of 1 $\mu$m for the K6V planets. Interestingly, because \ce{O3} production is diminished by about an order of magnitude for the K6V planets compared to the planets orbiting the Sun, the \ce{O3} Chappuis band centered near 0.6 $\mu$m is not apparent for the K6V planets. The UV \ce{O3} Hartley-Huggins band is still visible in all spectra for $\lambda$ $<$ 0.3 $\mu$m, and this band is notably not saturated for the Case 1 K6V planet. For planets with even lower \ce{O2} amounts than we simulate here, such as atmospheres with $\leq$ 0.1$\%$ PAL \ce{O2} possibly representative of the mid-Proterozoic \citep{Planavsky2014}, \ce{O2} itself will be extremely difficult to observe. Therefore, access to UV wavelengths will be particularly important for observing \ce{O3} and establishing the presence of oxygen in such planets' atmospheres \citep{Schwieterman2018_O3}.

\section{DISCUSSION}
\label{sec:discussion}
The exoplanet revolution has already surprised us with the discovery of worlds not represented in our solar system (e.g. hot Jupiters and super Earths) and the knowledge that there are entire systems of small planets on our cosmic doorstep \citep[e.g. the TRAPPIST-1 seven-planet system, orbiting an M8V star at 12 pc, ][]{Gillon2017}. By observational necessity, most of the current and near-future observations of exoplanets focus on M dwarfs, but these planets' prospects for habitability may be imperiled by high stellar activity levels and a lengthy super-luminous pre-main sequence phase. To maximize our chances of discovering habitable worlds and life elsewhere, we must seek also observations of temperate terrestrial planets orbiting ``Sun-like'' (i.e. F, G, and K) stars as emphasized in the recent \textit{Exoplanet Science Strategy} report \citep{ExoSciStrategy}. New facilities beyond those current planned would be required to accomplish these observations. 

\begin{figure*}[htb]
\begin{center}
\includegraphics[width=160mm,scale=1.0]{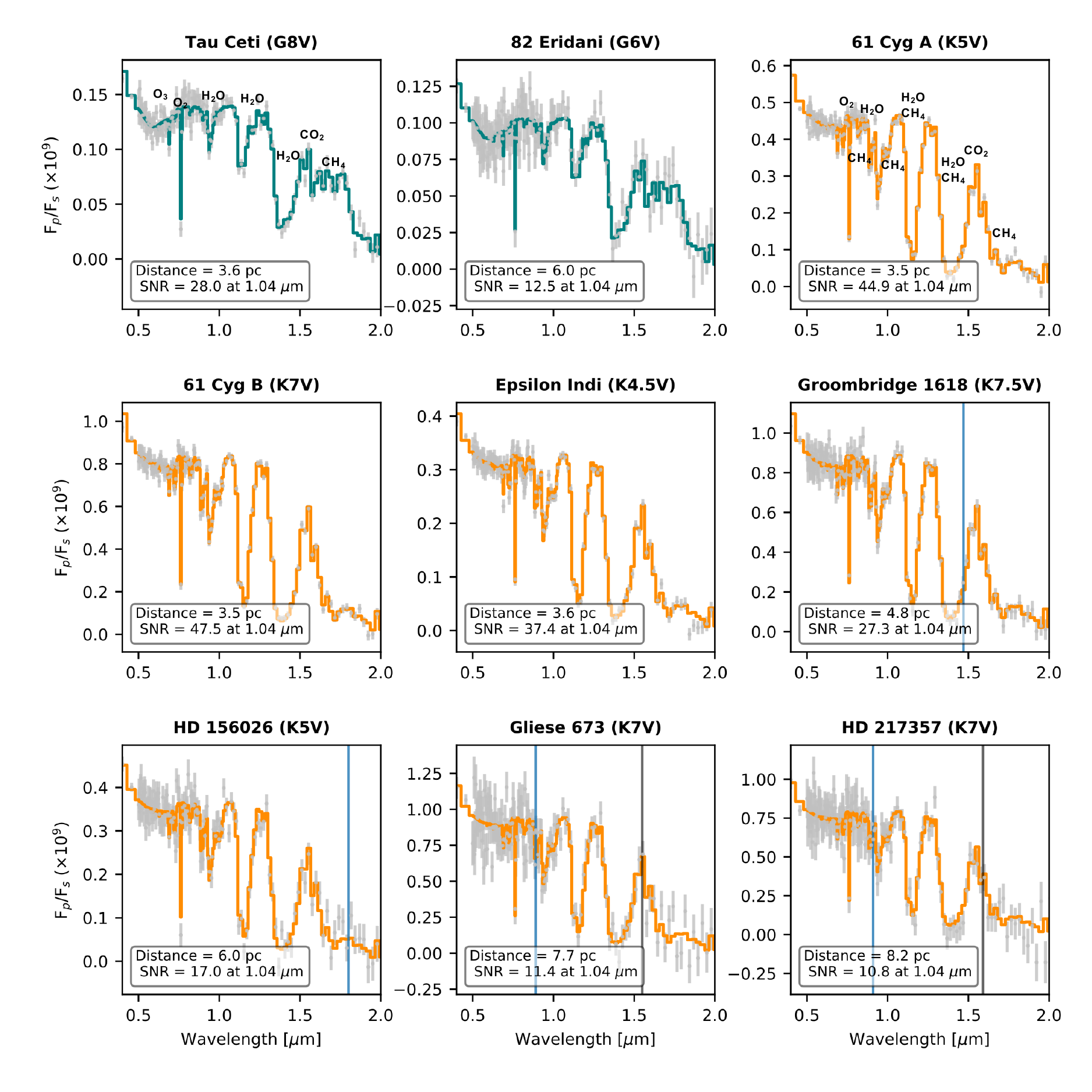}
\caption{Simulated observations of a Case 3 planet orbiting several nearby K dwarf stars (orange spectra) that might be targeted with future exoplanet direct imaging facilities.  Two nearby G dwarfs are shown for comparison (teal spectra). Important spectral features are labeled. Spectra are the same as those shown in Figure 3 for the Sun and the K6V star. Planets around all stars are placed at orbital distances where they receive 0.7 $\times$ Earth's isolation. Points with error bars (grey) simulate observations with a LUVOIR-A (15 m) telescope for 50 hours of integration time per coronagraph bandpass. Vertical blue and dark gray lines show the longest wavelength that can be observed for IWA = 3.5$\lambda$/D, and IWA = 2$\lambda$/D, respectively. }  
\end{center}
\label{fig:fig4}
\end{figure*}

As we have shown, K dwarfs, especially later stars, offer advantages over G dwarfs in the search for biosignatures because these stars' UV spectra allow for a longer photochemical lifetime of methane in oxygenated atmospheres, thus increasing the likelihood of detecting this disequilibrium biosignature gas pair. Future observatories could target a number of nearby mid-to-late K dwarfs, including the K6V star whose spectrum we used here (HD 85512, 11.6 pc), as well as: 61 Cyg A/B \citep[K5V/K7V, 3.5 pc; note the 61 Cyg binaries orbit each other with a period of 544 years,][]{Brocksopp2002}, Epsilon Indi (K4.5V, 3.6 pc), Groombridge 1618 (K7.5V, 4.8 pc), HD 156026 (K5V, 6.0 pc), Gliese 673 (K7V, 7.7 pc), HD 217357 (K7V, 8.2 pc), and HD 151288 (K7.5V, 9.8 pc). K dwarf planets orbiting these stars and others might be found by the PLAnetary Transits and Oscillations of stars (PLATO) satellite, which is expected to find K dwarf HZ planets; anticipated yields for PLATO of ``small'' planets (R $<$ 2 Earth radii) in the HZ of ``Sun-like'' stars range from less than 10 up to 280, depending on estimates of the fraction of stars with Earth-like planets.

One challenge that K dwarfs present for direct observations is that their HZ planets will be on orbits with smaller semi-major axes compared to planets orbiting G dwarfs. This means that planets orbiting K dwarfs are more vulnerable to falling inside the IWA of future observatories. The IWA denotes the smallest planet-star separation at which a planet can be resolved and will affect the ability of any direct imaging telescope, including LUVOIR and HabEx, to observe exoplanets.

HabEx is considering designs that include a starshade, which may offer a small enough IWA to observe nearby K dwarf HZ planets. For starshades, the IWA is proportional to the radius of the starshade and inversely proportional to the starshade-telescope separation distance. The HabEx 4-m telescope starshade concept is baselined to have an IWA of 60 mas \citep{HabEx2018} and would be able to observe 0.3 - 1 $\mu$m simultaneously. Longer wavelengths out to 1.8 $\mu$m could be accessed by repositioning the starshade closer to the telescope, but this will sacrifice IWA (IWA$_{NIR}$ = 108 mas). With the HabEx starshade, a planet orbiting at 0.42 AU from a K6V star, such as the one that we simulate here, could be observed at 0.3 - 1 $\mu$m out to about 7 pc (16 pc for an Earth-equivalent planet around a G2V star) and could be observed to 1.8 $\mu$m out to about 4 pc (10 pc for a planet around a G2V star). Note, however, that the conservative HZ for a K6V star extends to 0.65 AU \citep{Kopparapu2013}, so planets at the outer edge of the HZ could be observed for 0.3 - 1 $\mu$m out to 10 pc and for 1.8 $\mu$m out to 6 pc (26 and 15 pc, respectively, for planets at 1.6 AU around a G2V star).

Both LUVOIR and HabEx are considering designs that include coronagraphs. For a coronagraph, the IWA is dependent on wavelength and inversely related to telescope diameter: IWA $= c\lambda/D$, where c is a small-valued constant of order unity, $\lambda$ is wavelength, and D is telescope diameter. LUVOIR is exploring 15-m (on-axis) and 8-m (off-axis) observatory designs (LUVOIR-A and -B, respectively).

Different types of coronagraphs offer different IWAs: for instance, the apodized pupil Lyot coronagraph \citep[APLC,][]{Zimmerman2016} is tolerant to resolved stellar diameters but has a relatively large IWA of $\sim3.5 \lambda / D$ and takes a throughput hit from the apodizer mask. The vector vortex coronagraph \citep[VVC,][]{Ruane2016, NDiaye2015} is more sensitive to resolved stars for a centrally obscured telescope aperture but offers a smaller IWA of $\sim2 \lambda / D$.  The phase-induced apodization \citep[PIAA,][]{Guyon2010} is a third type of coronagraph that performs better for segmented apertures and offers an IWA of $\sim3 \lambda / D$. LUVOIR is exploring carrying the APLC, VVC, and/or PIAA coronagraphs on board \citep{LUVOIR2018}. HabEx is basedlined to carry a VVC coronagraph. 

There is a moderately strong \ce{CH4} band near 1 $\mu$m for methane-rich atmospheres such as our Case 3 planets, which might be observable for nearby K dwarfs. For our standard planet at 0.42 AU from a K6V star, a coronagraph with IWA  = $3.5 \lambda / D$ (APLC, only being explored by LUVOIR) could observe to 1 $\mu$m for planets at distances up to 9.5 and 5 pc for LUVOIR-A and -B, respectively (20 and 11 pc for planets around a G2V star). For a planet at the outer edge of the K6V HZ, these distance change to 13 and 7 pc for LUVOIR-A and -B (33 and 18 pc around a G2V star). Using a coronagraph with IWA = $2 \lambda / D$ (VVC, being explored by LUVOIR and HabEx), one can observe our standard K6V planet at 1 $\mu$m for distances up to 15, 8, and 4 pc for LUVOIR-A, -B, and HabEx (35, 19, and 10 pc around a G2V star). These distances change to 23, 12, and 6 pc for planets at the outer edge of the K6V HZ (55, 30, and 15 pc for planets around a G2V star, although these distances may be too far to obtain good signal).  

Despite the potential challenges of observing K dwarf planets outside the IWA of possible future observatories, these stars offer the major advantage of higher planet-star contrast compared to planets orbiting G dwarf stars. Thus, their spectra can be obtained in shorter integration times. A K6V star, for instance, is only about a tenth as bright as a G2V star (Table \ref{tab:tab1}). As a comparison, a LUVOIR-A telescope observing for 50 hours can obtain a signal-to-noise ratio (SNR) of 9.2 near 1 $\mu$m for a planet orbiting a solar analog at 7 pc. This increases by more than a factor of 2 to SNR = 20 for a planet orbiting a K6V star at 7 pc. At the LUVOIR-A visible resolution, R = 140, \citet{Feng2018} shows that SNR = 10 at V band can provide detection (i.e. a peaked posterior distribution) of \ce{O2, O3, H2O}, surface pressure, and planetary radius for modern Earth, while SNR = 15 does this and provides constraints (i.e. a peaked posterior distribution with 1-$\sigma$ width $<$ an order of magnitude) on \ce{O3}, surface pressure, and radius (note that Feng et al. do not discuss \ce{CH4}).

Figure \ref{fig:fig4} shows simulated 50 hour (per coronagraph bandpass) LUVOIR-A observations for Case 3 planets orbiting a variety of stars in the nearby stellar neighborhood that may be targeted by future habitable planet and biosignature searches. Two nearby G dwarfs, Tau Ceti and 82 Eridani, are included for comparison. The G dwarfs use the Case 3 atmosphere generated for the Sun, and the K dwarfs use the same for the K6V star presented in Section \ref{sec:results}. The planets orbiting the K dwarfs offer better SNR than the equivalent planets orbiting G dwarfs at similar distances due to improved planet-star contrast, and they may show stronger methane features in an \ce{O2}-rich atmosphere as highlighted here.  An IWA of $2\lambda / D$ (dark gray vertical line) is sufficient to allow access to strong \ce{O2}, \ce{H2O}, \ce{CH4}, and \ce{CO2} features for all K dwarfs shown. For the farthest stars shown, an IWA of $3.5\lambda / D$ (blue vertical line) will not allow access to \ce{CO2} features, and \ce{CH4} could only be detected via weaker visible wavelength bands for planets with sufficiently high \ce{CH4} enrichment such as these Case 3 worlds. 

These simulations suggest that nearby late K dwarfs such as 61 Cyg A, and 61 Cyg B, Epsilon Indi, Groombridge 1618, and HD 156026 may be particularly excellent targets for biosignature searches on exoplanets. In addition to the ``K dwarf advantage'' for biosignatures, these stars can offer access to a wide range of wavelengths for habitable zone planets even with IWA constraints. 61 Cyg A, and 61 Cyg B, Epsilon Indi, Groombridge 1618 provide higher or comparable SNR to Tau Ceti, the closest G dwarf other than the Sun and Proxima Centauri A. In particular, 61 Cyg A and 61 Cyg B, which are at a similar distance as Tau Ceti (3.6 pc), offer SNR that is 1.6 - 1.7 times better in the same integration time.  HD 156026 is at a similar distance as 82 Eridani (6 pc), and it offers 1.4 times better SNR compared to this G6V star.

We have shown that a sufficiently small IWA enables excellent characterization of nearby K dwarf HZ planets, so one of the most important technological innovations that could improve observations of nearby habitable K dwarf planets are observatories with small IWAs. This would provide access to redder wavelengths and/or planets orbiting more distant stars. An observatory like HabEx or LUVOIR would not launch until the 2030s or 2040s, so there is considerable time for maturation of promising coronagraph and starshade technologies. For instance, the visible nulling coronagraph (VNC) under development offers an excellent IWA ($2 \lambda / D$) with relatively high throughput due to its lack of apodizer mask, but its optical complexity is high. Technical development of the VNC is ongoing \citep{Hicks2016}.

\section{CONCLUSIONS}
\label{sec:conclusions}

The discovery of life on another planet would be a watershed moment in the history of science, with implications that would ripple throughout all of society. However, capturing and correctly interpreting the sparse stream of photons from distant inhabited exoplanets will be a formidable, awe-inspiring task even with powerful future telescopes. 

Oxygen and methane are important gases to seek in future biosignature searches because together they indicate an atmosphere in chemical disequilibrium and are a powerful indicator of life. Previous studies have shown that the photochemical lifetime of methane in an oxygenated atmosphere is longer around M dwarfs compared to G dwarfs. However, the habitability of M dwarf planets may be endangered by high levels of stellar activity and a prolonged super-luminous pre-main sequence phase. Here, we have explored how K dwarf photochemistry can produce simultaneously observable \ce{O2} and \ce{CH4} spectral features, finding that later K dwarfs may generate an order of magnitude more \ce{CH4} compared to equivalent planets around Solar-type stars. Because K dwarfs offer a better planet-star contrast ratio compared to G dwarfs, shorter observing times are needed achieve a given signal-to-noise ratio. Particularly nearby mid-to-late K dwarfs such as 61 Cyg A/B, Epsilon Indi, Groombridge 1618, and HD 156026 may be especially good targets for future biosignature searches on exoplanets. The practical requirements for observing exoplanets in the HZs of K dwarfs should therefore be carefully considered when planning for possible future exoplanet observatories.

\acknowledgments
{This work was performed as part of the NASA Astrobiology Institute's Virtual Planetary Laboratory, supported by the National Aeronautics and Space Administration through the NASA Astrobiology Institute under solicitation NNH12ZDA002C and Cooperative Agreement Number NNA13AA93A, and by the NASA Nexus for Exoplanet System Science (NExSS) research coordination network Grant 80NSSC18K0829. We thank the Goddard Space Flight Center Sellers Exoplanet Environments Collaboration (SEEC) for support, which is funded by the NASA Planetary Science Division's Internal Scientist Funding Model (ISFM).  Any opinions, findings, and conclusions or recommendations expressed in this material are those of the author(s) and do not necessarily reflect the views of NASA. The author thanks the reviewer for helpful comments that improved the quality of this manuscript.}

\emph{Software}: Atmos \citep{Arney2016}, SMART \citep{Meadows1996}, Coronagraph Noise Model \citep{Robinson2016}.

\bibliography{bibs}

\end{document}